\documentclass{ws-ijmpc}
\usepackage[latin1,utf8]{inputenc}
\usepackage[super,compress]{cite}
\usepackage{graphicx}
\usepackage{epsfig}
\usepackage{gnuplottex}
\usepackage[usenames]{color}
\usepackage{hyperref}
\hypersetup{colorlinks,
urlcolor=blue,
citecolor=blue,
linkcolor=blue}

\begin{document}

\markboth{Tiago Moy da Silva \& Américo T. Bernardes}
{Ripples and Grains Segregation on Unpaved Road}

\catchline{}{}{}{}{}

\title{Ripples and Grains Segregation on Unpaved Road}

\author{Tiago Moy da Silva\footnote{Present address: Departamento de Física, CCEN, Universidade Federal da Paraíba, 58059-900, João Pessoa, PB, Brazil.} \hyperlink{first}{$^{,\dagger}$} and Américo T. Bernardes\hyperlink{first}{$^{\S}$}}

\address{Departamento de Física,\\
  Universidade Federal de Ouro Preto,\\
  Ouro Preto, MG, 35400-000, Brazil\\
  \hypertarget{first}{$^{\dagger}$}tiagomoy@aluno.ufop.edu.br\\
\hypertarget{first}{$^{\S}$}atb@iceb.ufop.br}

\maketitle

\begin{history}
\received{Day Month Year}
\revised{Day Month Year}
\end{history}

\begin{abstract}
Ripples or corrugations are common phenomena observed in unpaved roads in less developed countries or regions. They cause several damages in vehicles leading to increased maintenance and product costs. In this paper we present a computational study about the so called washboard roads. Also we study grain segregation on unpaved roads. Our simulations have been performed by the Discrete Element Method (DEM). In our model, the grains are regarded as soft disks. The grains are subjected to a gravitational field and both translational and rotational movements are allowed. The results show that wheels of different sizes, weights and moving with different velocities can change corrugations amplitude and wavelength. Our results also show that some wavelength values are related to specific wheel's speed intervals. Segregation has been studied in roads formed by three distinct grains diameter distribution. We observed that the phenomenon is more evident for higher grain size dispersion.

\end{abstract}

\ccode{PACS Nos.: 81.05.Rm, 45.70.Mg, 02.70.-c}

\section{Introduction}

When driving in rural areas in developing countries, one observes that ripples or corrugations are a common phenomenon, leading those roads to be known as washboard roads. This effect appears due to the traffic of vehicles on roads covered by granular materials (sand or gravel, for instance) causing discomfort to users and damages to the vehicles \cite{0}. This is a common problem in rural regions and poor neighborhoods in developing cities and can affect the harvest flow between the production regions and the consumers, thus causing production losses and increased product prices. Hence, the study of the basic mechanisms of corrugations formation remains important. More than describing the phenomenon, it is also important to study how to minimize it.

Granular materials are systems formed by a collection of solids and macroscopic particles that interact with their neighbors through repulsive and dissipative contact force. It is assumed that the grains must be bigger than 1 micrometer, otherwise thermal agitation would be perceived \cite{9}. Some examples are sand, cement and rice. Discussion on granular materials main features can be found in some reviews, like in Refs. \cite{9,10,RevModPhys.71.435}. In the last decades were published several works about granular matter, for instance, pattern formation in granular systems (see Refs. \cite{6,7,8,8.5,umbanhowar1996localized}) and segregation, which is the most important phenomenon studied in the granular framework (see Refs. \cite{11,12,13,14,thornton2012modeling,15}).

There are several works about washboard road, such as \cite{1,2,3,OZAKI201511}. But our attention is focused on papers that follow the perspective of granular materials, like those of Taberlet \textit{et al.}\cite{4} or Bitbol \textit{et al.}\cite{5}. In both works a computational and experimental study were performed with a rolling wheel without any kind of traction or torque applied in the wheel and an inclined blade plow. Among other things, they show that segregation and compaction do not influence in the wave patterns formation and they study the instability in terms of critical velocity and the Froud number. It was obtained that the mass of the wheel affects the wave pattern, but the wheels diameter does not change the pattern. The wave pattern independence on the diameter of the wheel was a fundamental picture for the mathematical model detailed in Ref. \refcite{4}. Furthermore, in Ref. \refcite{5} it was considered variables as the axis in which the wheel was attached, which is a feature of the experimental apparatus. This model is described in therms of critical velocity; in the transition regime between flat and rippled road.

Hewitt \textit{et al.} \cite{hewitt} performed a study about the ripples formation in granular and fluid after the only one passage of the plow. Percier \textit{et al.}\cite{percier1} investigated via experiments and simulations the drag and lift forces on a inclined plow as well as they performed numerical calculations of stress and velocity fields. In other work Percier \textit{et al.}\cite{percier} study a stability analysis based on measurement of lift force on a plow. And, most recently, Srimahachota \textit{et al.}\cite{PhysRevE.96.062904} performed an experimental investigation about the influence of a harmonic oscillator attached in a cylinder which is dragged (it does not roll) on a granular surface. 

Considering the importance of both effects: ripples formation and segregation; we performed a DEM study to simulate an unpaved road. Here, we modeled a granular bed with polydisperse two-dimensional grains. The wheel is a large disk that rolls on the road. We then observe the results by changing some physic and geometrical parameters. Differently from Taberlet \textit{et al.}\cite{4} or Bitbol \textit{et al.}\cite{5}, where the wheel's axis moves with a constant velocity, in our model the wheel's motion is created by the application of a constant torque. As we have cited above, the results obtained in these two papers were based in a specific experimental apparatus. Then, in the present paper we shall explore possibilities with physical quantities not yet considered. Another difference from their models is that we performed simulations with inhomogeneous roads, i.e., we used grains with different diameters for each simulation. It allows us to better understand the segregation phenomenon. Remembering that Taberlet \textit{et al.}\cite{4} and Bitbol \textit{et al.}\cite{5} showed that segregation is not important in the ripples formation; however it occurs in real roads (see Ref. \refcite{0}), which justifies our study about segregation on unpaved roads.

In Sec. \ref{sec:level2} we introduce the force model in granular matter and the methods to characterize ripples and segregation. In Sec. \ref{sec:level3} we show the results of the numerical simulations and the respective discussion. In Sec. \ref{sec:level4} we present a summary and suggestions for further works.


\section{\label{sec:level2}Methodology and Details}
\subsection{Model}

\begin{figure}
\begin{center}
\includegraphics[width=0.6\textwidth]{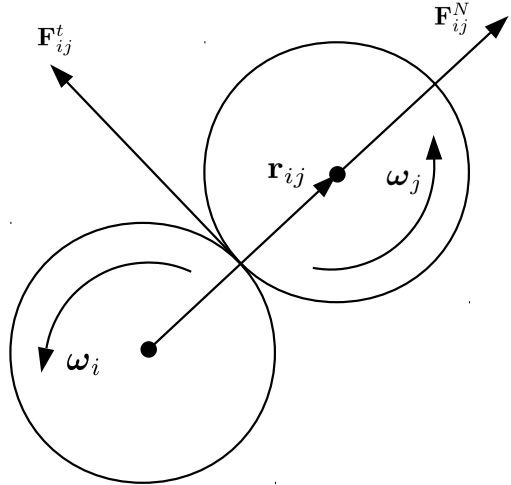}
\caption{Vectors diagram for the contact without compression of two disks.}
\label{figforca}
\end{center}
\end{figure}

In this two-dimensional model the grains are regarded as soft-disks which are deposited on the bottom of a simulation box. The wheel is a large disk whose initial position is chosen so that it is placed above the road, contacting it. Dry grains interact by contact force. In this model, this contact force is given by the sum of the normal and tangential components (Fig. \ref{figforca}). This description of the interaction between grains has been used successfully in several papers on granular material, as for example in Refs. \refcite{15} and \refcite{16}. For a pair of grains the force is given by
\begin{equation}
\textbf{F}_{ij}= \left\{ \begin{array}{ll}
\textbf{F}_{ij}^{N}+\textbf{F}_{ij}^{t} & \textrm{ if $\alpha>0$}\\
0 & \textrm{ otherwise ,}\\
\end{array} \right.
\label{forca}
\end{equation}
where $\alpha = R_i+R_j-|\textbf{r}_{ij}|$ is the overlap between two grains, $R_i$ and $R_j$ are, respectively, $i$ and $j$ the radius, and $|\textbf{r}_{ij}|$ is the disks centers distance. The normal force is\cite{16}
\begin{equation}
\textbf{F}_{ij}^N= \left[\kappa \alpha - \gamma(\hat{\textbf{r}}_{ij} \cdot \textbf{v}_{ij}) \right]\hat{\textbf{r}}_{ij},
\label{norm}
\end{equation}
where $\kappa$ is the elastic constant, $\hat{\textbf{r}}_{ij}$ is the unit vector in the normal direction, $\gamma$ is the damping coefficient and $\textbf{v}_{ij}$ is the relative velocity between disks.

The tangential component is a sliding friction force given by \cite{15}
\begin{equation}
\textbf{F}_{ij}^t=-\min \left(\gamma^t|\textbf{v}_{ij}^t|,\mu|\textbf{F}_{ij}^N|\right)\hat{\textbf{v}}_{ij}^t,
\label{tang}
\end{equation}
where $\gamma^t$ is the sliding friction constant (also called shear damping coefficient, see Ref. \refcite{HAFF1986239}), $\mu$ is the static friction coefficient, $|\textbf{F}_{ij}^N|$ is the normal force modulus, given by Eq. (\ref{norm}), and $\textbf{v}_{ij}^t$ is the relative transverse velocity at the contact point, as given in Refs. \refcite{16} and \refcite{20}.

The length unit $L$ is equivalent to size of a grain with a diameter equal to 1; the force $F$ and time $T$ are given in units of $\kappa L$ and $\sqrt{M/ \kappa}$, respectively. $M$ is the mass unit. The time step used in the simulations was $\tau=0.001$ $T$ and the equations of motion for each pair of particles are integrated using the leapfrog algorithm, as in Ref. \refcite{20}. At this point it is appropriate to define a cycle, which is an evolution of the whole system in one time step $\tau$. If $N_c$ is the number of the cycles, the total time of simulation is $\tau N_c$. Thus, the graph of any physical quantity plotted as function of cycles evolves also in time. The values of the elastic constants used in the simulations were: $\kappa=40,000$ $F/L$, $\gamma=10$ $FT/L$. These values lead to the following restitution coefficients\cite{17}, $e_l=0.94597$ for collision between two grains with $d_i=1$ $L$ and $e_s=0.97803$ for two grains with $d_i=0.4$ $L$ colliding. 

The tangential force given by Eq. (\ref{tang}) has a maximum value that obeys Coulomb's friction law, $\textbf{F}_{ij_{\textrm{max}}}^t= \mu|\textbf{F}_{ij}^N|$ and it does not allow static friction because it depends on the relative velocity between the grains. For $\textbf{v}_{ij}^t=0$, the tangential force vanishes. So, in Eq. (\ref{forca}) only the normal component will be considered. This feature in the model causes a slow dissolution of the ripples, as reported in Ref. \refcite{poschel2005computational}. By using the values of the parameters $\gamma^t=100$ $FT/L$ and $\mu=0.6$, the ripples reduce less than $1$ $L$ per each round of the wheel and therefore it does not affect the wave pattern. The contact force between wheel and grain is calculated in the same way described above for the contact force between grains. 

The whole system is under action of a vertical gravitational field, $\textbf{g}=-g\hat{\textbf{j}}$. Here we have adopted $g=5$ $L/T^2$. This is an arbitrary definition, which reflects the basic parameters used to define the wheels and grains.

The road (simulation box) has periodic boundary conditions (PBC) in the horizontal direction: if some particle or the wheel leave the simulation box on the right side, they return to the box through the left side and vice versa \cite{18}.

\subsection{Segregation}

Segregation means that grains of same size agglomerate. So, we may consider that they form clusters. As it has been done in previous works \cite{19} in order to observe the segregation in the road we have calculated the radial distribution function [RDF, indicate by $g(\textbf{r})$] defined by \cite{20,21}:
\begin{equation}
g(\textbf{r})=\frac{2}{\rho N} \left \langle \sum_{i<j}^{N}\delta(\textbf{r}-\textbf{r}_{ij}) \right \rangle,
\label{rdf}
\end{equation} 
where $\rho$ is the density (per unity area) and $N$ is the number of particles. The sum in the angular brackets is related to local density around an arbitrary particle and the delta function assumes the values
\begin{equation}
\delta(x-x_i)= \left\{ \begin{array}{ll}
\infty & \textrm{ if center of grain $i$ is located at $x$}\\
0 & \textrm{ if center of grain $i$ is not at $x$.}\\
\end{array} \right.
\end{equation}

The two-dimensional discrete form of $g(r)$ is calculated by using a histogram of distance between pairs of grains \cite{doi:10.1142/S0129183110015725}
\begin{equation}
g(r)=\frac{Ah_n}{\pi N^2r\Delta r},
\label{rdfhistogram}
\end{equation} 
where $A$ is the area, $h_n$ is the number of grain pair $(i,j)$ within the interval $r-\Delta r/2 \leq r_{ij}\leq r+\Delta r/2$ and $\Delta r$ is a small distance. Since initially the grains are randomly distributed inside the road, and thus we suppose they are uniformly distributed, segregation should occur after several passages of the wheel. We apply the RDF in grains of same diameter in order to known if the local density increases in relation to the initial system.

As an alternative method, we have calculated the mean height of the largest or smallest particles in the beginning and at the end of the simulation. If the largest particles move to the top of the road, we should observe that the mean height at the simulations end should be greater than in the beginning.
 
\subsection{Discrete Fourier Transform (DFT)}

Waves are described by parameters such as amplitude and wavelength. Giving the road's shape after corrugations have been formed, we can calculate their values. This was made by splitting the simulation box in $\mathcal{N}$ columns, all of them with a width equal to $1$ $L$. In each of those columns we determine the center of mass coordinate of the highest grain. Therefore, repeating from first to the last column and summing to each center of mass height its particle radius we obtain a sequence of numbers that represent the height $h_x$ in function of the position $x$, i.e., the road's profile.

The sinusoidal shape of the road $h_x$ was transformed, via DFT, for period's space. The DFT is defined by \cite{22}:
\begin{equation}
H_k \equiv \sum_{x=0}^{\mathcal{N}-1} h_x \exp(2 \pi ixk/\mathcal{N}),
\label{dfteq}
\end{equation}
where $k$ is the variable of period and $\mathcal{N}$ is the number of columns in the simulation box. The wave pattern obtained is not regular so our main objective is to identify the period of the largest amplitude mode. As we can see below, the transformation shows a principal mode and some noise, due to the many random factors, like imperfections on the road. In the results discussed in the next section, only the value of the principal mode will be shown.


\section{\label{sec:level3}Results and discussion}

\subsection{Simulation details}

There are several parameters that can influence the final results in our corrugations study. So, they have to be changed in order to observe this influence. As we have pointed out above, three types of grain diameters distributions have been used. In the first distribution, the particles diameters are given by four values ($25\%$ of the quantity of grains for each value of $d_i$), $d_i=1.0$, $0.8$, $0.6$ and $0.4$ $L$; in the second, there are also four values (once again, $25\%$ for each value of $d_i$), $d_i=1.0$, $0.9$, $0.8$ and $0.7$ $L$; and in the last distribution the diameters values are uniformly distributed between $0.80\leq d_i \leq 1.20$ $L$. Grain masses are given by $m_i=\pi \rho_p d_i^2/4$ where $\rho_p$ is the density of each particle. 

\begin{figure}
\centering
  \includegraphics[width=0.9\textwidth]{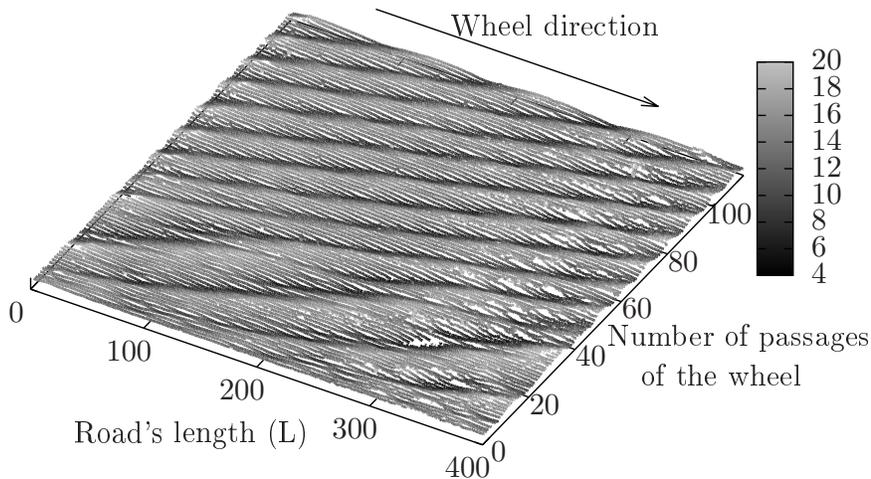}
\caption{The figure shows the evolution of the waves according to the wheel passages. The displacement of the waves occurs in the same direction of the movement of the wheel. The road is formed by diameter distribution 1 with 10,000 grains, the diameter wheel has $50$ $L$ and mass $200$ $M$.} 
\label{fig-3}
\end{figure}

In order to know the influence of granular bed height on the final result, we performed simulations in systems with $10,000$, $15,000$ and $30,000$ particles, and road's length equal to $400$ and $1,500$ $L$.

Three different wheels have been used in our simulations. The first wheel has $D_1=30$ $L$ and $M_1=100$ $M$, the second $D_2=50$ $L$ and $M_2=200$ $M$ and the third wheel has $D_3=100$ $L$ and $M_3=600$ $M$. Due to the fact that the applied torque has fixed value for each specific simulation, wheel's velocity is not controlled. This velocity depends on the road's changing shape as well as on the intensity of torque. Thus, for each simulation we calculate the mean velocity, $\langle V_{\textrm{x}} \rangle$, and all the results are presented as a function of it. All simulations are performed during $2\times10^7$ cycles.

\begin{figure}
\centering
\includegraphics[width=0.23\textwidth,angle=270]{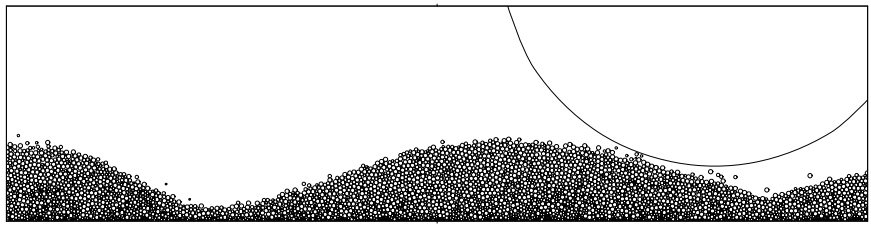}
\caption{The picture shows a wheel of diameter equal to $100$ $L$ when it surmounts a ripple. The wheel moves from left to right and the road is formed by grains of the distribution 1.} 
\label{pic}
\end{figure}

Our simulations are two-dimensionals, however we expect that the results are qualitatively satisfactory in both ripples and segregation. Generally, two-dimensional simulations in granular systems are used to study some aspect in which it has a difficult experimental setup, as for instance to investigate properties of internal structure of a dense granular system, as in Ref. \refcite{4}. On the other hand, two-dimensional simulations make it difficult to perform a quantitative analysis when compared to experimental measures. Due to the low number of experimental publications with well controlled parameters there are not enough data for a direct comparison between our numerical results and some experimental results. But, to further works the natural units of simulation can be easily converted to any system of units as needed.

\subsection{Characterization of the ripples}

The changes in the several quantities in the simulation modify the ripples amplitude and wavelength. We have performed simulations by combining each road type, given by a specific grain diameter distribution, and different values of applied torque. Each torque leads to a different value of the average velocity $\langle V_{\textrm{x}} \rangle$. The values of $\langle H_k \rangle$ and $\langle \lambda \rangle$ have been obtained in ten measures, the first one was made at the $19,100,000$ cycle, the second at the $19,200,000$ cycle and so on. The tenth measure at the $2\times10^7$ cycle.

\begin{figure}
  \centering
\includegraphics[width=0.9\textwidth]{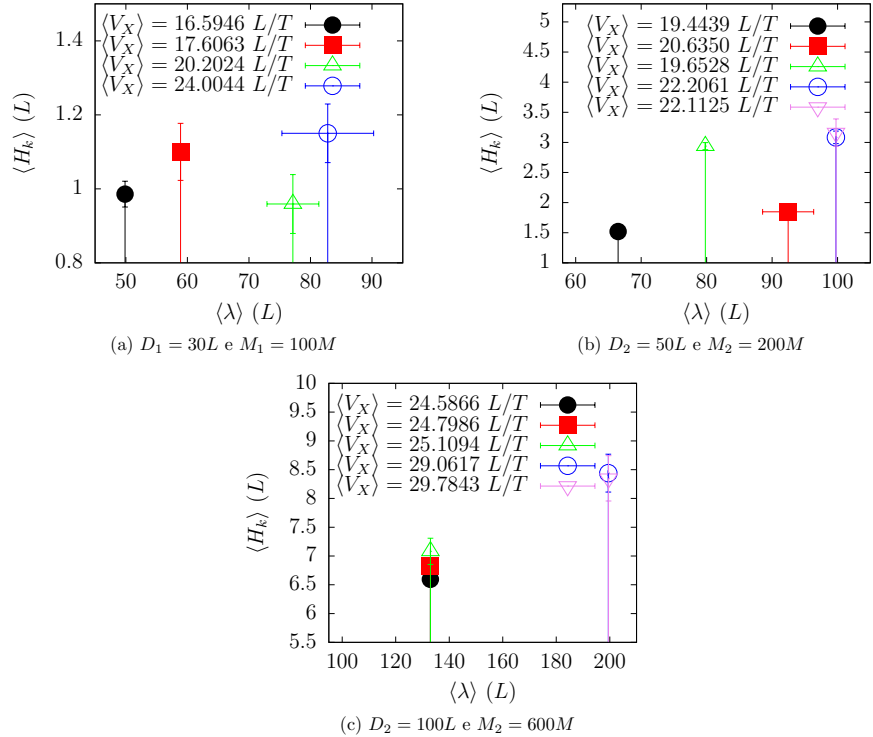}
\hypertarget{dft-a}{\caption{Results of DFT for different wheels on road formed by distribution of diameter type 2. Simulations have been performed with 10,000 grains in the box with 400 $L$ long. We observed that by increasing the torque increased wavelength. For each simulation, the mean velocity was calculated and showed in the inner box. See text for better discussion.}}
\end{figure}

Figure \ref{fig-3} shows the evolution of the ripples for each passage of the wheel on the road. In the first 10 passages we observe a region in which the road is approximately flat (in fact there are small waves but it does not show well defined pattern and are imperceptible in the scale of the figure). Larger waves appear after an unstable interval. After that, between 40 and 50 passages, the waves split and more waves with minor wavelengths can be observed. After 50 passages the pattern becomes well defined without major changes. We can also observe that the ripples move in the same direction of the wheel.

If the wheel moves slowly the granular bed remains flat while for much faster wheels the ripples do not show well defined patterns. Thus, well-defined ripples emerge only for an interval of velocities, as we can see in Fig. \ref{pic}. It was observed that oscillations amplitude and wavelength depend either on the type of wheel or on the road. In all cases, larger values of the $\langle V_{\textrm{x}} \rangle$ generated longer wavelengths and, in some cases, different speeds can produce the same value of $\lambda$. This possibly indicates that there is a speed range of the wheel for which, when reaching a maximum, smaller waves melt leading to the generation of larger ripples; in other words, each speed range of the wheel generates a specific wavelength. This feature cannot be observed in the present work, since in these simulations there is no such speed control. However, it is proposed as a suggestion for future works.

\begin{figure}
  \centering
\includegraphics[width=0.9\textwidth]{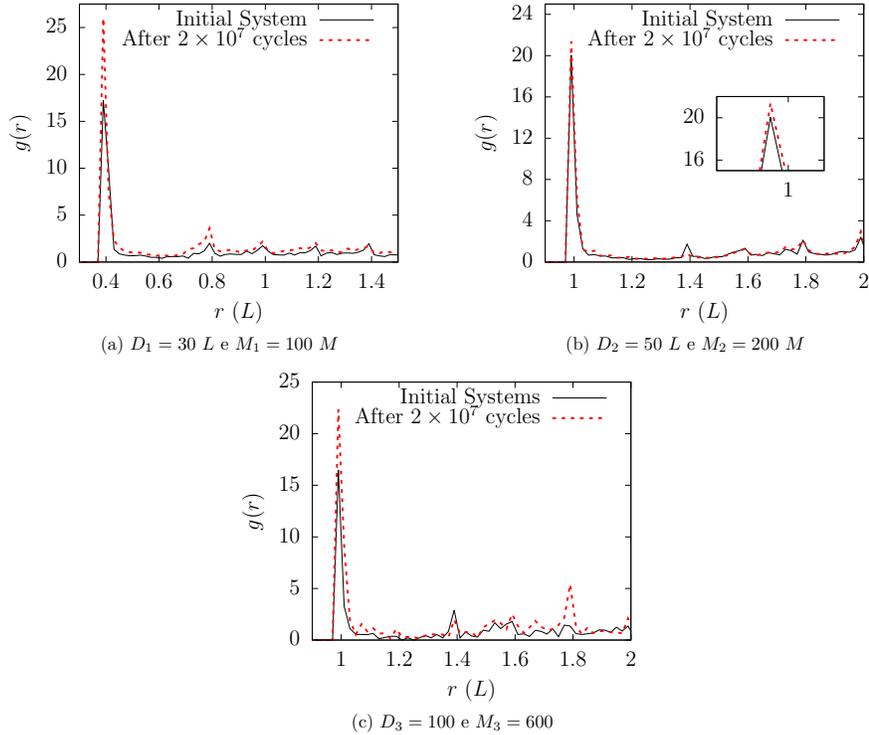}
\hypertarget{rdfp}{\caption{In (a) the RDF was applied to the grains of diameter $d=0.4$ $L$. The graph (b) show the RDF applied to the grains of diameter $d=1.0$ $L$ and the inset is a zoom to detail the height of the peaks. The RDF graph (c) show the RDF for $d=1.0$ $L$ in the case in which the diameter distribution has different quantities of each diameters. In this case has 650 grain of diameter $d=1.0$ $L$.}}
\end{figure}

\begin{figure}
\centering
\includegraphics[width=0.9\textwidth]{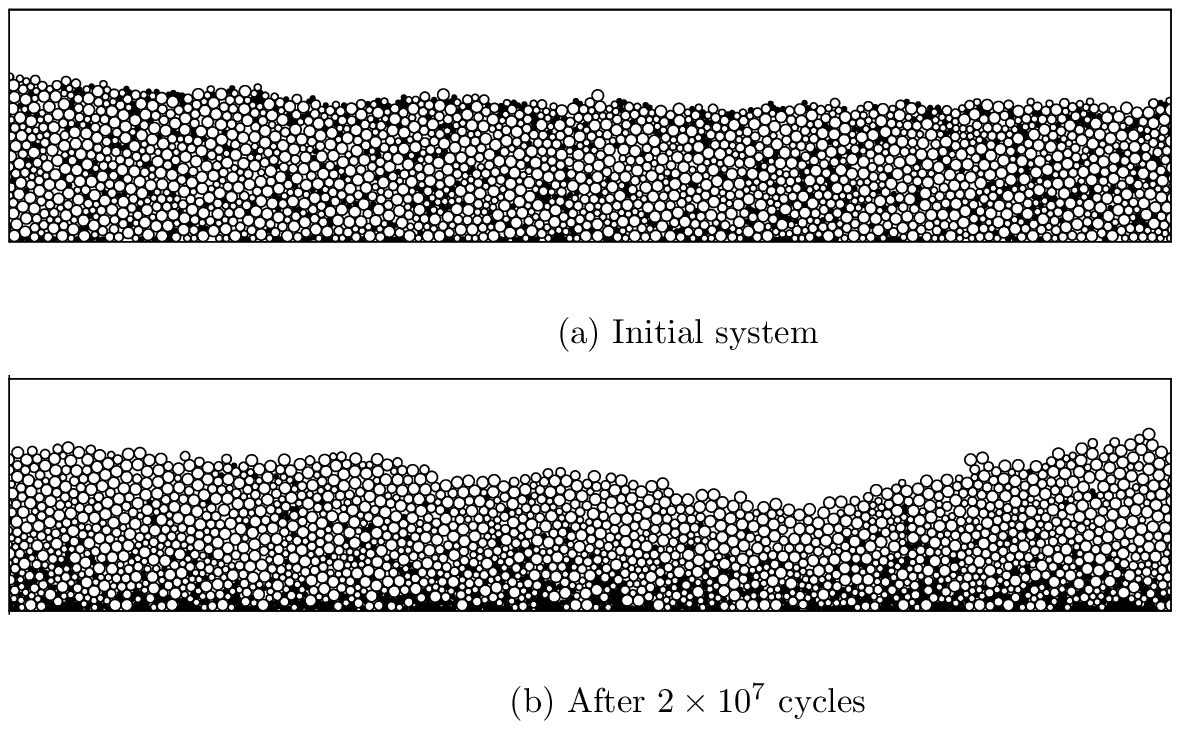}
\caption{Enlarged picture of road formed by 10000 grains with distribution 1. On top, initial configuration and at the bottom, the final configuration.  The small particles (in black) agglomerates at the bottom and the average height of large grains increased. This indicate that occurs segregation of the large and small particles.}
\label{segreg}
\end{figure}

Simulations with three different wheels led to the appearance of ripples with different characteristics. Larger wheels produce higher amplitudes and wavelengths for the three cases of the grain diameters. Fig. \hyperlink{dft-a}{4} shows DFT plots obtained for diameter distribution type 2. In the first case, the plot \hyperlink{dft-a}{4(a)} shows undulations with a mean wavelength between 50 and 82 and low amplitudes. The difference between the highest and lowest amplitude on the roads is less than 1. In the other results obtained with the same grain distribution [see Figs. \hyperlink{dft-a}{4(b)} and \hyperlink{dft-a}{4(c)}] it can be seen that the wavelength and amplitudes increase for larger wheels and the amplitudes did not show a pattern of growth. The behavior described above is qualitatively equal when applied in the diameter distributions of type 1 and 3. This effect occurs due to the combination of the wheel's rotation, due to the torque, and the contact between granular bed and the wheel. If the wheel's diameter is large, more particles are thrown in the opposite direction to the movement of the wheel, thus increasing the number of grains on the crest of the ripples and hence increasing the amplitude. This is a factor to the changing of the wave pattern observed in Fig. \ref{pic} when compared with references previously cited \cite{4,5}. There is also the impact when the wheel hits the wave valley, then causing that grains are expelled from the valley to the highest part of the undulation. In summary, the torque can modifying the characteristic of pattern wave according the wheel type or velocity.

Comparing the roads of different diameter distributions, we observed that distributions 1 and 2 produced larger and well stabilized undulations when simulating smaller wheels. For larger wheels, were obtained well defined patterns in all the diameter distributions. 

The change in the road height has also been tested, formed by diameter distribution 1, for the two bigger wheels. In both cases there were almost identical results for roads with 10,000 and 15,000 grains. For road with 30,000 grains we observed that the wave amplitude changed but the wavelength obtained was about the same as for the other lower roads. The length of the road was also changed to 1,500 with 30,000 particles, but the ripples remained with the same characteristics.

\subsection{Segregation}

As described above, segregation was studied through the characterization of the radial distribution function - RDF - for both larger and smaller grains of each distribution. For the first distribution the behavior of particles with diameters $d_i=1.0$ and $0.4$ have been analyzed; for distribution 2, the RDF was calculated for particles with diameters $d_i=1.0$ and $0.7$ $L$ and for type 3 the application was taken between the following intervals $0.8\leq d_i\leq 0.9$  and $1.1\leq d_i\leq 1.2$ $L$, since the particles diameters belong to a continuum range. Figure \hyperlink{rdfp}{5(a)} shows the RDF results obtained for grains with $d= 0.4$ $L$ of the distribution 1 for wheel 1. As expected, the first peak of the RDF occurs at a distance equal to $0.4$ $L$ and, by comparing the initial and final configurations, we observe that for this size of grains the segregation occurs at the end of simulation. In contrast, when the RDF was calculated for particles of diameter $d=1.0$ $L$ the plots do not indicate segregation. This can be seen in the Fig. \hyperlink{rdfp}{5(b)}, where the segregation was not identified [see inset of Fig. \hyperlink{rdfp}{5(b)}]. Similar results were obtained for wheels of type 2 and 3. 

\begin{figure}
\centering
\includegraphics[width=0.9\textwidth]{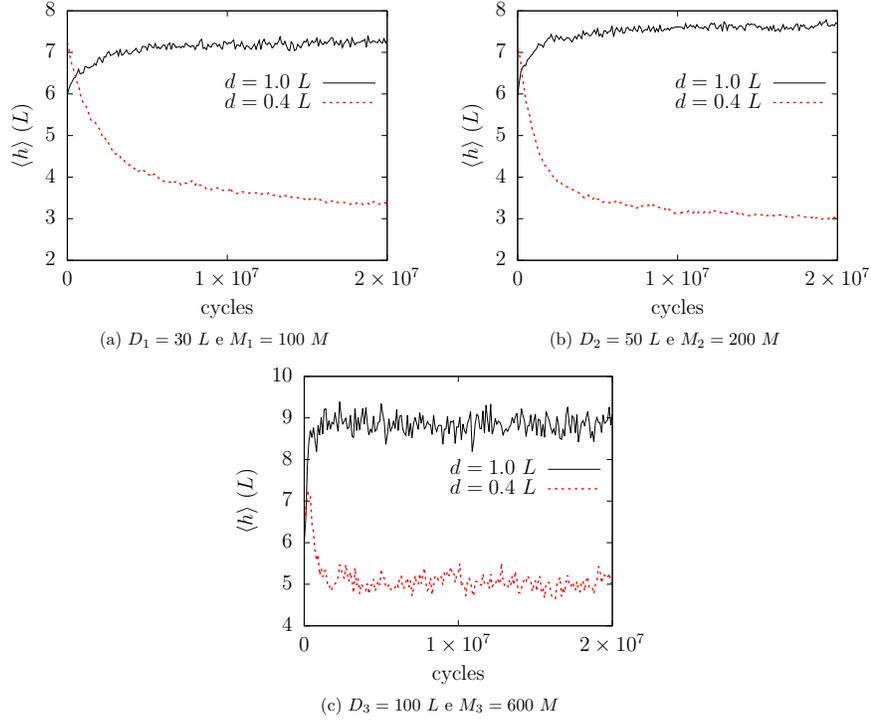}
\caption{Average height versus cycles of the grains of diameter $d=1.0$ $L$ and $d=0.4$ $L$ for three different wheels.} 
\label{heig123}
\end{figure}

As discussed above, there is a case in which the segregation does not occur. But as one can see in Fig. \ref{segreg} larger grains are at the top of the road, and it was not identified by the RDF calculation. It happens because we have so many larger particles in this simulation that they filled much of the road's volume (in this 2d case, area) and the segregation cannot be well observed through RDF calculation. Then a simulation was performed with a road formed by a distribution that has 650 grains with $d=1$, 2850 with $d=0.8$, 5000 with $d=0.6$ and 6500 grains with $d=0.4$ $L$. The result correspondent to this diameter distribution is showed in the Fig. \hyperlink{rdfp}{5(c)}, where it is possible to verify the occurrence of segregation in grains of diameter equal to 1.

For a more detailed analysis we plotted three graphs of the mean height versus cycles for the three types of wheels evolving on roads formed by diameter distribution 1, as shown in Fig. \ref{heig123}. In the three cases there is an increase of $\langle h \rangle$ of the largest grains and a decrease when the diameter is small. For longer periods the values tend to a constant value. The effect of different types of wheels in the graphs is perceived at the time in which the grains reach the maximum or minimum values of $\langle h \rangle$. Wheel type 3 leads to $\langle h \rangle$ reaching the maximum faster than the other types, while for wheel type 1 we observed the slowest process.

Segregation in the roads formed by distributions 2 and 3 has also been observed, but in both cases it occurs with less intensity, due to the fact that the values of the diameter have a small difference in size and low dispersion. This mechanism is similar to the BNE (Brazil nut effect), shown in Ref. \refcite{13}, where the grains migrate through the vacancy among the grains when the granular packing is shaken. When the wheel passes on the road it causes a perturbation which shakes the grains of the granular bed. Since the distribution 1 has greater dispersion in diameter, the segregation is more intense. 


\section{\label{sec:level4}Conclusions}

In this work we present a computational study about two common effects in granular materials: ripples and segregation; both observed in an unpaved road. The simulations were performed by changing some variables in the road and in the wheel. The analysis of the wave patterns showed that velocity, diameter and mass of the wheel change the characteristics of the ripples because of the torque applied to the wheel. Our results are in contrast to Refs. \refcite{4} and \refcite{5}, in which it was observed that only the mass of the wheel change the ripples pattern. The torque, applied to the wheel, plays a role of traction and is responsible for these changes. In some cases the same wavelength was obtained for different mean velocities indicating that it is possible that there are intervals of velocities related to a given wavelength. 

Regarding the segregation, the simulations showed that it is more intense in the diameter distribution 1 due to the vacancies in the road which are larger than in the other ones. The RDF was not sufficient to evaluate the biggest grains segregation. Thus, we calculated the average height of those grains of diameter equal to 1. The increase of the average height indicates that there was segregation. As previously stated, segregation is not essential for ripples formation, however this study was justify by the fact that in real roads the segregation almost always occurs together with the ripples formation.

The computational study about corrugations and segregation on unpaved roads allows us a better comprehension of these very common effects, from the perspective of granular materials. As in other areas of this field, we note the absence of a closed theory capable of explaining the main observed features in the dynamics of granular materials. For further research, we can modify the values of the constants, as for example the spring constant and the coefficient of friction, model the torque in order to better control the average velocity, create a mathematical theory based on wheel-road interaction and perform more experiments, since there are still few experimental publications on this topic.

\section*{Acknowledgments}

Tiago Moy da Silva acknowledges support by a UFOP grant. The authors acknowledge partial financial support from CNPq, CAPES, and FAPEMIG (Brazilian agencies).



\end{document}